\documentclass[aps, prb, superscriptaddress, reprint, twocolumn, amsmath, amssymb, notitlepage]{revtex4-1}
\usepackage{graphicx}
\usepackage{amsmath,amssymb}
\usepackage{color}
\usepackage{physics}
\usepackage[utf8]{inputenc}

\begin{document}

\title{Optical signatures of electron--phonon decoupling due to strong light--matter interactions}

\author{Emil V. Denning}
\affiliation{Department of Photonics Engineering, Technical University of Denmark, 2800 Kgs. Lyngby, Denmark}
\affiliation{NanoPhoton - Center for Nanophotonics, Technical University of Denmark, Ørsteds Plads 345A, DK-2800 Kgs. Lyngby, Denmark}
\email{emvo@fotonik.dtu.dk}

\author{Matias Bundgaard-Nielsen}
\affiliation{Department of Photonics Engineering, Technical University of Denmark, 2800 Kgs. Lyngby, Denmark}
\affiliation{NanoPhoton - Center for Nanophotonics, Technical University of Denmark, Ørsteds Plads 345A, DK-2800 Kgs. Lyngby, Denmark}

\author{Jesper M\o rk}
\affiliation{Department of Photonics Engineering, Technical University of Denmark, 2800 Kgs. Lyngby, Denmark}
\affiliation{NanoPhoton - Center for Nanophotonics, Technical University of Denmark, Ørsteds Plads 345A, DK-2800 Kgs. Lyngby, Denmark}

\begin{abstract}
Phonon interactions in solid-state photonics systems cause intrinsic quantum decoherence and often present the limiting factor in emerging quantum technology. 
Due to recent developments in nanophotonics, exciton--cavity structures with very strong light--matter coupling rates can be fabricated. We show that in such structures, a new regime emerges, where the decoherence is completely suppressed due to decoupling of the dominant phonon process. 
Using a numerically exact tensor network approach, we perform calculations in this non-perturbative, non-Markovian dynamical regime. Here, we identify a strategy for reaching near-unity photon indistinguishability and also discover an interesting phonon-dressing of the exciton--cavity polaritons in the high-$Q$ regime, leading to multiple phonon sidebands when the light--matter interaction is sufficiently strong.
\end{abstract}

\maketitle

\section{Introduction}
The development of scalable solid-state quantum technology is challenged by lattice vibrations, i.e. phonons, which even at zero temperatature deteriorates the quantum coherence~\cite{fu2009observation,iles2017phonon}. The interaction of electrons and phonons thus leads to remarkable features in the optical emission spectrum, such as broad spectral sidebands and incoherent scattering~\cite{brash2019light,tran2016quantum,christiansen2017phonon,doherty2013nitrogen,selig2016excitonic}. This is detrimental to the optical coherence and important to circumvent for applications in quantum technology. It also presents an open quantum system with rich physics, operating in a regime of pronounced non-Markovian dynamics~\cite{carmele2019non}.

Recent developments in nanophotonics have opened up the possibility of creating dielectric nanocavities with deep subwavelength confinement of light~\cite{choi2017self}, leading to light--matter interaction strengths otherwise far beyond reach in dielectrics~\cite{hu2018experimental,wang2018maximizing}. Moreover, experiments have demonstrated very high coupling strengths between a plasmonic nanocavity and two-dimensional transition metal dichalcogenides~\cite{wang2016coherent,kleemann2017strong,stuhrenberg2018strong,han2018rabi,geisler2019single,qin2020revealing} or a single dye molecule~\cite{chikkaraddy2016single}. These developments open the door to a new regime of nanophotonic electron--phonon interactions, where the light--matter coupling rate is comparable to or larger than the dominating phonon frequencies in the environment. Previous theoretical studies in the context of quantum chemistry have shown that decoupling of electronic and nuclear dynamics in chemical reactions can occur in this regime~\cite{galego2015cavity,herrera2016cavity,galego2016suppressing}.
In this paper, we theoretically study the impact of electron-phonon decoupling on light emission from exciton--cavity systems and identify the fundamental requirements for complete elimination of phonon signatures in the generated light. 
We consider a generic system consisting of an exciton mode coupled to a single quantized cavity mode and a continuum of phonon modes. Here, the comparability of phononic and optical time scales makes calculations of the dynamical properties highly challenging and has demanded extensive development of non-perturbative and non-Markovian theoretical methods~\cite{hornecker2017influence,morreau2020phonon,morreau2019phonon,vagov2011real,kaer2010non,kaer2013microscopic}. In this work, we have implemented a numerically exact and computationally efficient tensor network formulation, which allows us to calculate two-time averages~\cite{jorgensen2019exploiting,strathearn2018efficient}, thus forming the basis for assessing optical emission properties. Furthermore, we make use of a variational polaron perturbation theory to derive analytical results that explain the dynamical decoupling process.

As an important example system, we consider a nanocavity containing a semiconductor quantum dot, which is coupled to the continuum of longitudinal acoustic phonon modes of the host lattice~\cite{wilson2002quantum,winger2009explanation,besombes2001acoustic,senellart2017high}. For this system, we calculate the emission spectrum and the photon indistinguishability, which is a useful and generic measure of the optical coherence~\cite{mandel1991coherence}.  We find that the interplay between the phonon cutoff frequeny (i.e. the dominating vibrational frequency scale in the environment), the light--matter coupling strength and the cavity decay rate determines the type of phonon decoupling process that can be observed. Specifically, the phonon signatures in the optical emission can be completely suppressed, when the nanocavity is in the low-$Q$ Purcell regime and the light--matter interaction strength exceeds the phonon cutoff frequency. This opens a new route towards realizing single-photon sources with near-unity photon indsitinguishability. Additionally, we predict a novel, interesting effect in the high-$Q$ limit, where each of the exciton polariton peaks in the spectrum is dressed with an individual phonon sideband, demonstrating non-perturbative dynamics, where polaritons and polarons occur at an equal footing. The observed decoupling effects can occur for any type of excitonic system and relies only on the general form of the exciton--phonon coupling.

\section{The electron--phonon decoupling regime}
Our analysis is based on a generic system consisting of a localised exciton state, $\ket{X}$, a cavity mode with annihilation operator $a$ and a vibrational environment with phonon annihilation operators $\{b_\mathbf{k}\}$. When an exciton is created, the equilibrium position of the ions of the lattice or molecule is displaced due to the electrostatic interaction. This leads to an exciton--phonon coupling described by the Hamiltonian~\cite{mahan2013many}
\begin{align}
H_{\rm ep} = \dyad{X}\sum_\mathbf{k} \hbar (g_{\mathbf{k}} b_\mathbf{k} +   g_{\mathbf{k}}^* b_\mathbf{k}^\dagger ),
\end{align}
where $\{g_{\mathbf{k}}\}$ are the exciton--phonon coupling strengths. The free evolution of the phonons is governed by the Hamiltonian $H_{\rm p}=\sum_\mathbf{k}\hbar\nu_\mathbf{k}b_\mathbf{k}^\dagger b_\mathbf{k}$, where $\nu_\mathbf{k}$ is the frequency of the phonon mode with momentum $\mathbf{k}$. Together with the coupling, $H_{\rm ep}$, this defines the phonon spectral density, $J(\nu) = \sum_\mathbf{k} \abs*{g_\mathbf{k}}^2\delta(\nu-\nu_\mathbf{k})$, which fully characterises the influence of the vibrational environment on the exciton. For any realistic physical system, this spectral density has a cutoff frequency, $\xi$, such that $J(\nu)\simeq 0$ for $\nu \gg \xi$. This cutoff frequency is related to the length scale of the exciton wavefunction and the properties of available phonon modes in the material~\cite{alkauskas2014first,nazir2016modelling,mahan2013many}. The evolution of the exciton--cavity system is governed by the Hamiltonian 
\begin{align}
H_{\rm s} = \hbar\omega_X\dyad{X} + \hbar\omega_c a^\dagger a +  \hbar g(\dyad{0}{X} a^\dagger + \dyad{X}{0} a),
\end{align}
where $\omega_X$ and $\omega_c$ are the resonance frequencies of the exciton and cavity, respectively, $g$ is the light--matter coupling strength and $\ket{0}$ is the electronic ground state. Furthermore, cavity losses with a rate $\kappa$, exciton losses with a rate $\gamma$, and exciton dephasing with a temperature-dependent rate $\gamma^*(T)$ are treated through the Lindblad formalism~\cite{breuer2002theory,lindblad1976generators} as Markovian effects~\cite{reigue2017probing,tighineanu2018phonon,muljarov2004dephasing}. To describe the optical emission properties of the system, we initialise it in the exciton state with zero photons in the cavity and calculate the spectral correlation function of the emitted photons as the system relaxes, $S(\omega,\omega')=\kappa\ev*{a^\dagger(\omega)a(\omega')} = \kappa\int_{-\infty}^\infty\dd{t}\int_{-\infty}^\infty\dd{t'}e^{-i(\omega t-\omega't')}\ev*{a^\dagger(t')a(t)}$. From this spectral function, we can calculate the emission spectrum as $S(\omega,\omega)$~\cite{steck2007quantum}. In addition, it provides access to the coherence properties of the emitted photons, for example their indistinguishability~\cite{kiraz2004quantum}, $\mathcal{I}=[\int\dd{\omega}S(\omega,\omega)]^{-2}\int\dd{\omega}\int\dd{\omega'}\abs{S(\omega,\omega')}^2$, which quantifies the interference visibility of two subsequently emitted photons. 

\begin{figure}
  \centering
  \includegraphics[width=\columnwidth]{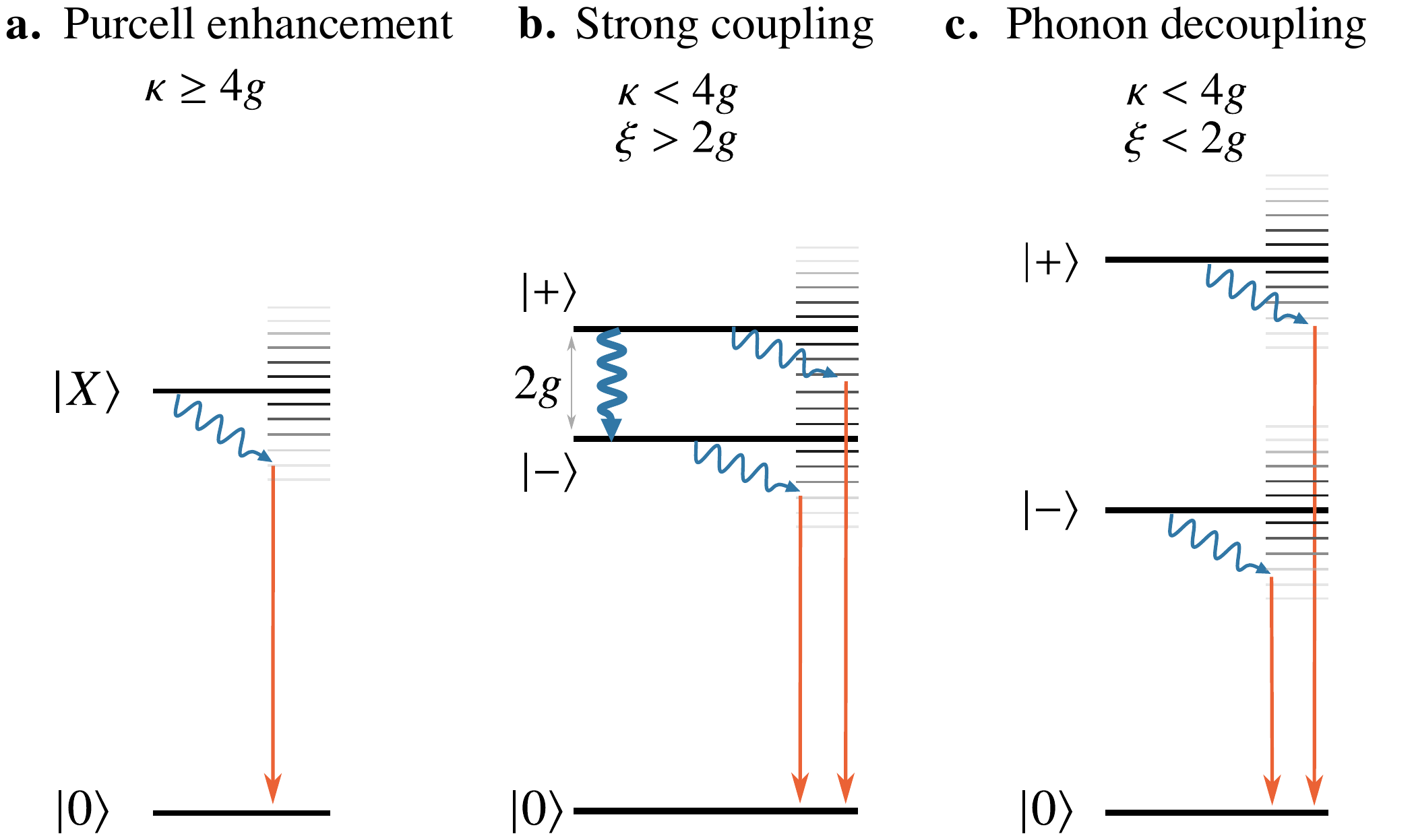}
  \caption{Illustration of phonon-mediated optical emission processes. {\bf a.} In the Purcell regime, the exciton decays and emits a photon (orange arrow). During this process, a phonon wavepacket (blue wiggly arrow) might be emitted or absorbed, resulting in a photon with lower or higher energy.  {\bf b.} In the strong light--matter coupling regime, a phonon wavepacket can be emitted either by relaxation from the upper polariton to the lower one (downwards wiggly arrow), or when one of the polaritons decays to the ground state. {\bf c.} In the phonon decoupling regime, where the polariton splitting, $2g$, exceeds the phonon cutoff frequency, $\xi$, the phonon sidebands on the two polaritons do not overlap and are hence spectrally resolved.}
  \label{fig:phonon-effects}
\end{figure}

There are three main parameter regimes of this system: In the Purcell regime (Fig.~\ref{fig:phonon-effects}a), attained when $4g<\kappa$ (in the limit where pure dephasing can be neglected), the exciton decays and emits a photon into the cavity with a rate of $\Gamma=4g^2/\kappa$. In this process, a phonon wavepacket may be emitted or absorbed, generating a broad sideband in the emission spectrum.
At low temperatures, $k_{\rm B}T\ll\hbar\xi$, the sideband is asymmetric and red-detuned from the zero-phonon line, reflecting that phonon emission dominates over phonon absorption~\cite{krummheuer2002theory}. In the strong coupling regime (Fig.~\ref{fig:phonon-effects}b), the coupling strength exceeds the decay, $4g>\kappa$, but is still well below the phonon cutoff frequency. Here, the exciton and cavity form hybrid polaritons, $\ket{\pm}=\ket{1,0}\pm\ket{0,X}$ (where $\ket{n,e}$ denotes a $n$-photon cavity state and electronic state $e\in\{0,X\}$) that are spectrally well-resolved and split by a frequency of $2g$. The dominating decoherence mechanism in this regime arises from a resonant transition from the upper polariton to the lower polariton under the emission of a phonon wavepacket with energy $\sim 2\hbar g$. If the temperature is sufficiently high to populate the phonon modes, the reverse process can also take place by phonon absorption. At low temperatures, the phonon emission process, $\ket{+}\rightarrow\ket{-}$, dominates, and a spectral polariton asymmetry can be observed, because photons are thus predominantly emitted from the lower polariton state~\cite{roy2015quantum,denning2020phonon, morreau2019phonon}. Since the polariton splitting is small compared to the phonon cutoff frequency, the sideband seen in the Purcell regime is not resolved into contributions from the two polaritons. 

Increasing the coupling strength further leads to a regime of phonon decoupling (Fig.~\ref{fig:phonon-effects}c), where $2g$ exceeds the phonon cutoff frequency. Due to this, there are no phonon modes with sufficiently high energy to drive polariton transitions, and this decoupling leads to a recovery of the quantum coherence. Additionally, the spectral symmetry between the polariton peaks is restored and the polaritons are now so far separated that the individual phonon-polariton sidebands are spectrally resolved. 

\begin{figure*}
  \centering
  \includegraphics[width=\textwidth]{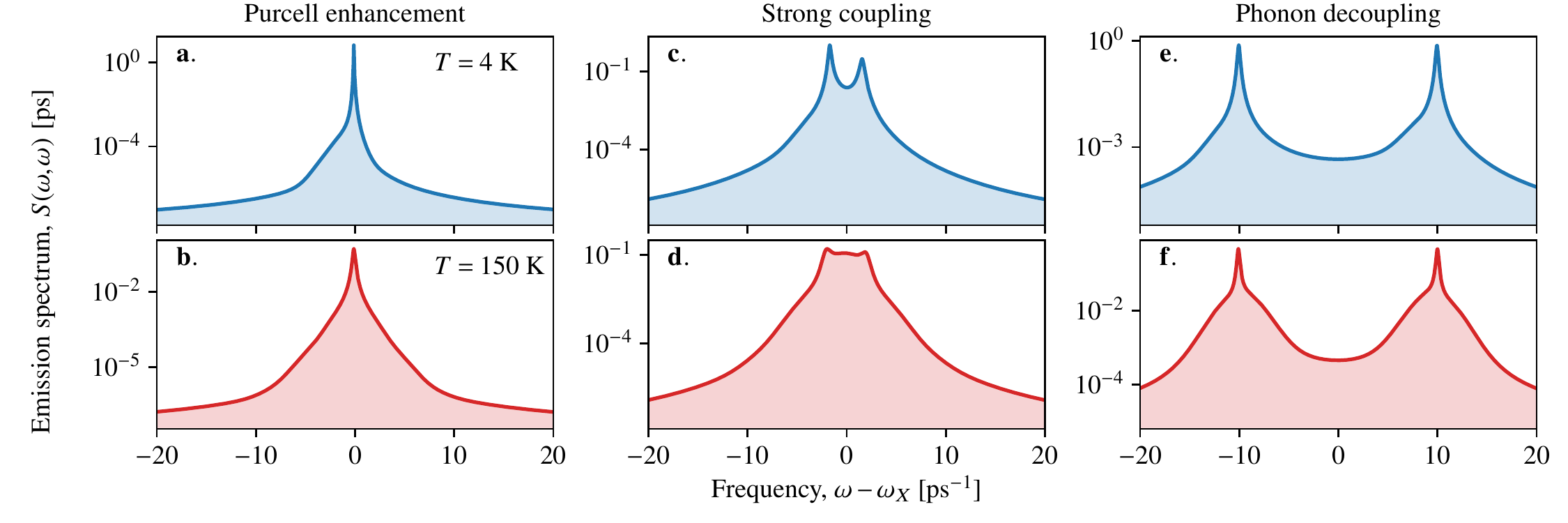}
  \caption{Optical emission spectra in of a quantum dot in a cavity in the {\bf a.-b.} Purcell regime $g=0.05\mathrm{\;ps^{-1}},\;\kappa =0.5\mathrm{\;ps^{-1}}$, {\bf c.-d.} strong coupling regime, $g=1.1\mathrm{\;ps^{-1}},\;\kappa =0.5\mathrm{\;ps^{-1}}$, and {\bf e.-f.} phonon decoupling regime, $g=10.0\mathrm{\;ps^{-1}},\;\kappa =0.5\mathrm{\;ps^{-1}}$. For the upper panels, the temperature is $T=4\mathrm{\;K}$, and for the lower panels, $T=150\mathrm{\;K}$. We have used an overall phonon coupling strength, $\alpha=0.025\mathrm{\;ps^2}$ and phonon cutoff frequency, $\xi = 2.23\mathrm{\;ps^{-1}}$, corresponding to a GaAs quantum dot of size 3 nm. The cavity and exciton were taken resonant, $\omega_c=\omega_X+R_\mathrm{v}$, and the exciton decay $\gamma=0.01\mathrm{\:ps^{-1}}$.}
  \label{fig:spectra}
\end{figure*}

Calculating the temporal correlation function entering $S(\omega,\omega')$ is a technically demanding task due to the non-Markovian interactions with the phonon environment. Our approach, based on a tensor-network representation of the phonon influence functional, is described in App.~\ref{sec:tens-netw-impl}. To illustrate the three different regimes in Fig.~\ref{fig:phonon-effects}, we use a semiconductor quantum dot in a nanocavity as an example. Here, the phonon cutoff frequency is typically on the order of a few $\mathrm{ps^{-1}}$~~~\cite{denning2020phonon}, and the spectral density is $J(\nu) = \alpha \nu^3\exp{-(\nu/\xi)^2}$, where $\alpha$ is an overall phonon coupling strength~\cite{nazir2016modelling}. The optical emission spectra for parameters corresponding to the three characteristic parameter regimes are shown in Fig.~\ref{fig:spectra}. The spectra in the upper panels are calculated for a temperature of $T=4\mathrm{\; K}$, and in the lower panels for $T=150\mathrm{\; K}$. 

In the Purcell regime (Fig.~\ref{fig:spectra}a-b), the spectrum exhibits a narrow zero-phonon line dressed by a broad phonon sideband, which is asymmetric in the low-temperature limit. In addition, thermal phonon scattering and dephasing broadens the zero-phonon line at higher temperatures. In the strong coupling regime (Fig.~\ref{fig:spectra}c-d), the polariton peaks are asymmetric at low temperature, and the polariton peaks are dressed by a single phonon sideband. In the regime of phonon decoupling (Fig.~\ref{fig:spectra}e-f), the polaritons are split beyond the phonon cutoff frequency, and thus the polaritons are dressed by spectrally resolved sidebands. Furthermore, the polariton symmetry in the spectrum is recovered.

Earlier theoretical work has identified a related phenomenon for an emitter driven with an external laser field, described by a semiclassical driving, instead of the quantised cavity mode considered here. Specifically, a re-appearance of Rabi oscillations were observed when the laser driving strength exceeds the phonon cutoff frequency~\cite{vagov2007nonmonotonic}. These findings were later supported by intuitive and accurate perturbation theories based on a variational polaron transformation~\cite{mccutcheon2011general,nazir2016modelling} that led to a deeper understanding of the system, and it has further been shown that such variational strategies could be combined with coupling to a quantised cavity mode~\cite{gomez2018solid}. Inspired by these approaches, we now proceed to develop a variational polaron theory for our system with a quantised cavity mode and use this theory to interpret our numerical results.

\section{Variational polaron theory}
\label{sec:vari-polar-theory}
The variational polaron formalism is based on a unitary transformation generated by the operator 
\begin{align}
V=\dyad{X}\sum_\mathbf{k}\frac{f_\mathbf{k}}{\nu_\mathbf{k}}(b_\mathbf{k}^\dagger - b_\mathbf{k}),
\end{align}
which transforms the Hamiltonian as $H_\mathrm{v} = e^{V}He^{-V}$. The parameters $f_\mathbf{k}$, which define the transformation, are then determined by minimizing the Feynman-Bogoliubov upper bound on the free energy~\cite{silbey1984variational,harris1985variational} (see App.\ref{sec:variational-minimisation}). 
In the variational frame, the system part of the Hamiltonian is given by
\begin{align}
\label{eq:H-s-v}
\begin{split}
H_{\rm s,v} &= \hbar(\omega_{X} + R_\mathrm{v})\dyad{X} + \hbar\omega_c a^\dagger a \\ &\hspace{1.5cm} +\hbar g_\mathrm{v}(\dyad{X}{0} a + \dyad{0}{X}a^\dagger),
\end{split}
\end{align}
with $R_\mathrm{v} = \sum_\mathbf{k} f_\mathbf{k}(f_\mathbf{k}-2g_\mathbf{k})/\nu_\mathbf{k}$ the variational renormalisation of the exciton transition frequency and $g_\mathrm{v}=gB_\mathrm{v}$ is the variationally renormalised light--matter coupling strength, with the renormalisation factor $B_\mathrm{v}=\ev{e^{\pm V}}= \exp[-\frac{1}{2}\sum_\mathbf{k}\frac{f_\mathbf{k}^2}{\nu_\mathbf{k}^2}\coth(\beta\hbar\nu_\mathbf{k}/2)]$, where $\beta=1/k_{\rm B}T$.

The effect of the variational polaron transformation is to dress the excitonic dipole operator by a vibrational displacement, such that $\dyad{0}{X}\rightarrow \dyad{0}{X}e^{-V}$. In this dressing, the displacement of the phonon mode with momentum $\mathbf{k}$ depends on the the relative magnitude of $g$ and $\nu_\mathbf{k}$; modes with $\nu_\mathbf{k}\ll 2g$ are effectively left undisplaced, and modes with $\nu_\mathbf{k}\gg 2g$ are displaced by $g_\mathbf{k}/\nu_\mathbf{k}$. In the intermediate regime, $\nu_\mathbf{k}\simeq 2g$, the modes are displaced between these two limits. This $\mathbf{k}$-dependent displacement reflects the ability of the phonon modes to follow the dynamics of the exciton and cavity. Thus, the variational theory predicts that phonon modes with frequencies below $2g$ are effectively decoupled from the exciton-cavity system, because they are too slow to follow the vacuum Rabi oscillations between the exciton and cavity. Since the cutoff frequency, $\xi$, sets the characteristic frequency scale for the phonon modes that interact with the exciton, the variational theory predicts that all the relevant phonon modes are decoupled when $2g\gtrsim \xi$. However, as we have seen in Fig.~\ref{fig:spectra}, this is not the full story: The polaritonic phonon sidebands that emerge in the decoupling regime are a manifestation of vibrational dressing of the polaritons, which persist even though $2g\gg \xi$.

An important characteristic that quantifies the transformation, is the variational renormalisation factor, $B_\mathrm{v}$, which depends on $g$ and takes a value between 0 and 1, such that $B_\mathrm{v}\simeq 1$ when $2g\gg\xi$ (see Fig.~\ref{fig:indist}a). The significance of $B_\mathrm{v}$ is two-fold: First, the light--matter interaction in the transformed Hamiltonian, $H_\mathrm{v}$, is renormalised as $g\rightarrow gB_\mathrm{v}$, meaning that the phonons reduce the effective coupling strength. Furthermore, when the exciton--cavity system is in the low-$Q$ Purcell regime, $4g<\kappa$, and $\kappa \gg \xi$, the probability of generating a phonon wavepacket jointly with the emission of a photon is given by $1-B_\mathrm{v}^2$, i.e. the phonon sideband constitutes a fraction of $1-B_\mathrm{v}^2$ of the total emission spectrum; in the limit $g\rightarrow 0$, $B_\mathrm{v}^2$ reduces to the Franck-Condon factor~\cite{iles2017phonon}. However, as shown in Fig.~\ref{fig:spectra}c-d, this branching ratio does not hold in the phonon decoupling regime, where the polariton peaks are dressed with a phonon sideband, even though $g$ is sufficiently large to ensure $B_\mathrm{v}\simeq 1$. Thus, the polaritonic phonon sidebands are a strongly non-perturbative effect that cannot be captured even by the variationally optimised perturbation theory.
In analogy with the coupling strength renormalisation, the variational transformation also shifts the exciton resonance by $R_\mathrm{v}=\sum_\mathbf{k}f_\mathbf{k}(f_\mathbf{k}-2g_\mathbf{k})/\nu_\mathbf{k}$. This effect is of minor importance, but needs to be taken into account when setting the cavity frequency to resonance with the exciton.

\section{Restoring the optical coherence}
\label{sec:restoring-coherence}

\begin{figure}
  \centering
  \includegraphics[width=\columnwidth]{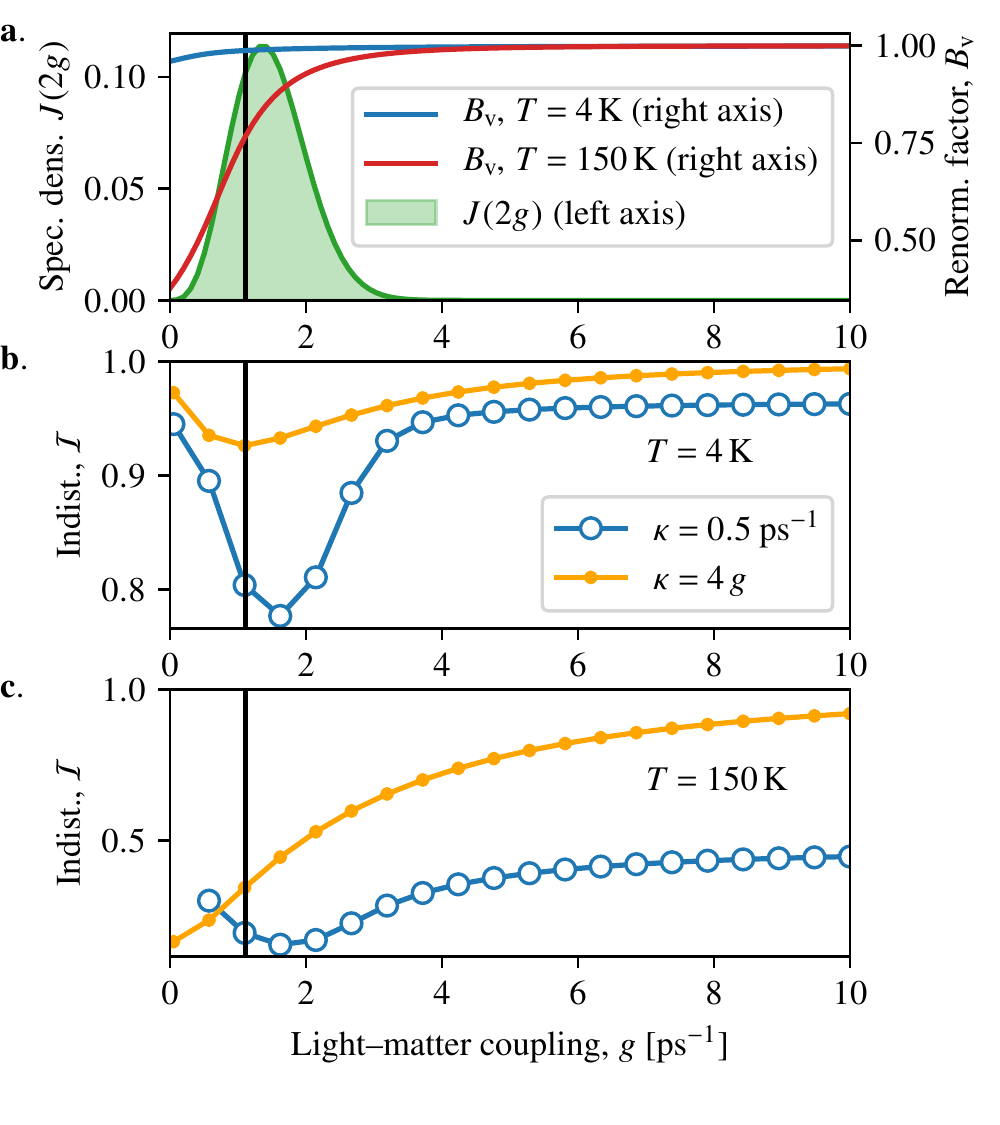}
  \caption{ {\bf a.} Phonon spectral density evaluated at $\nu=2g$ (green, left axis) and variational renormalisation factor, $B_\mathrm{v}$, (right axis) at $T=\mathrm{\; 4K}$ (blue) and $T=150\mathrm{\; K}$ (red) as a function of the light--matter coupling strength, $g$. The phonon cutoff frequency is indicated as $\xi/2$ by a solid black line. {\bf b.} Photon indistinguishability as a function of light--matter coupling strength, $g$ for fixed cavity decay, $\kappa=0.5\mathrm{\;ps^{-1}}$ (blue line and open circles) and cavity decay rate pinned to the coupling strength, $\kappa=4g$ (orange line and dots) at $T=4\mathrm{\;K}$. {\bf c.} Same as in b., but at $T=150\mathrm{\; K}$. The line signatures are the same as in panel b. The exciton decay is $\gamma=0.01\mathrm{ps^{-1}}$ for all calculations.}
  \label{fig:indist}
\end{figure}

To investigate the overall influence of the phonons in the decoupling regime, the photon indistinguishability is shown in Fig.~\ref{fig:indist}b as a function of $g$. The blue line with open circles signify a configuration with fixed cavity decay rate, corresponding to the blue spectra in the upper panels of Fig.~\ref{fig:spectra}. Here, it is clearly seen that the impact of the phonon environment is most significant when $J(2g)$ is maximal, meaning that the scattering process from the upper polariton to the lower is resonantly enhanced, and the photon emission process is exposed to strong decoherence. However, when $2g$ exceeds the cutoff frequency, the indistinguishability converges to $\sim 0.95$, due to the persistent polariton phonon sidebands. Alternatively, the orange line with dots shows the indistinguishability in a Purcell-configuration, where $\kappa$ is pinned at $4g$, ensuring that the system never enters the strong coupling regime. Here, the phonon sideband can be completely eliminated, when the zero-phonon line broadens sufficiently to absorb the entire sideband. 

The difference between the polariton and Purcell regimes becomes even more pronounced in the high-temperature limit (Fig.~\ref{fig:indist}c), where the sideband is more dominating. Due to thermal phonon population, the exciton dephasing here is stronger, meaning that the increase in indistinguishability with light--matter coupling strength is slower than for the low-temperature case. It is noteworthy that even at this high temperature, it is possible to achieve phonon decoupling and thus near-unity indistinguishability.

\begin{figure}
  \centering
  \includegraphics[width=\columnwidth]{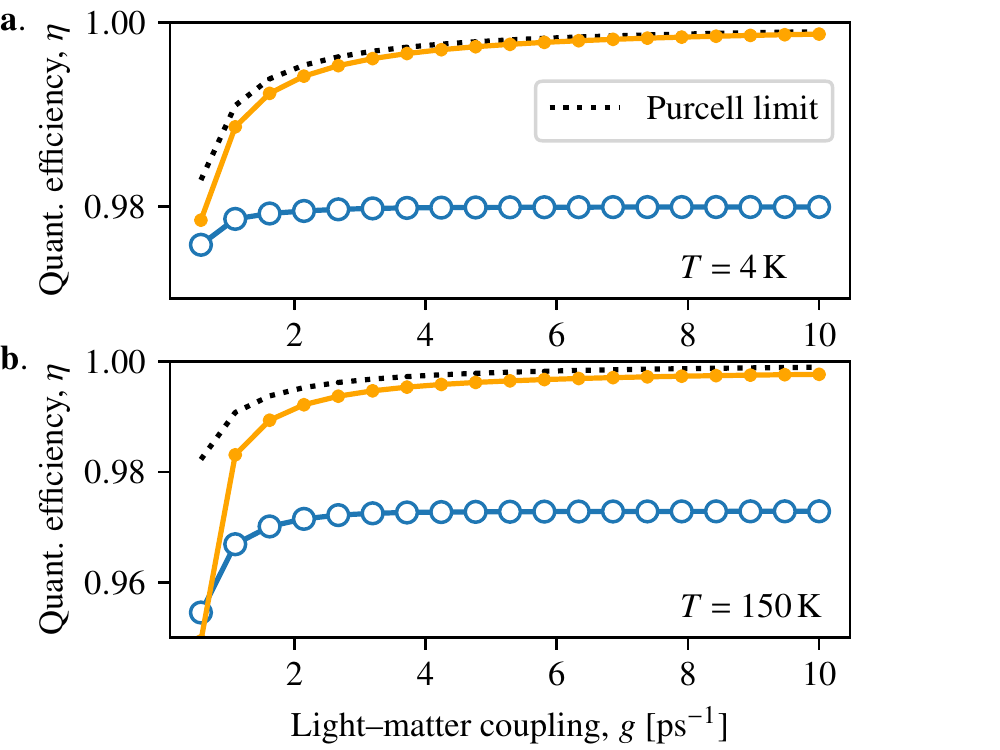}
  \caption{Quantum efficiency of photon emission, $\eta$, as a function of exciton--cavity coupling strength, $g$. The parameters are the same as in Fig.~\ref{fig:indist}.}
  \label{fig:efficiency}
\end{figure}
In addition to the photon coherence, the quantum efficiency is another important feature of interest in the contex of single-photon sources~\cite{gregersen2013modeling}. The quantum efficiency is the probability that an exciton is successfully converted into a photon in the detection channel. In our case, the cavity mode is the relevant detection channel, and the efficiency can be calculated as~\cite{iles2017phonon}
\begin{align}
 \eta = \kappa\int_0^\infty\dd{t}\ev*{a^\dagger(t)a(t)}.
\end{align}
This efficiency is plotted in Fig.~\ref{fig:efficiency} for $T=4\mathrm{\;K}$ (panel a) and $T=150\mathrm{\;K}$ (panel b) for the same parameters as the indistinguishability in Fig.~\ref{fig:indist}. In the Purcell regime, the decay of the exciton into cavity mode can be approximated as a Markovian process with the rate $\Gamma=4g^2/[\kappa+\gamma^*(T)]$~\cite{andrews2015photonics}. This process competes with the exciton losses due to other processes (most notably spontaneous emission into non-cavity modes), with the rate $\gamma$. Thus, the efficiency approaches $\eta_{\rm Purcell}=\Gamma/(\Gamma+\gamma)$ (shown with black dotted lines), and thus asymptotically approaches unity with increasing light--matter coupling. In the configuration with fixed cavity decay, the emission of photons into the detection channel is limited by $\kappa$ and the quantum efficiency therefore converges to a sub-unity value as the coupling strength increases.


\begin{figure}
  \centering
  \includegraphics[width=\columnwidth]{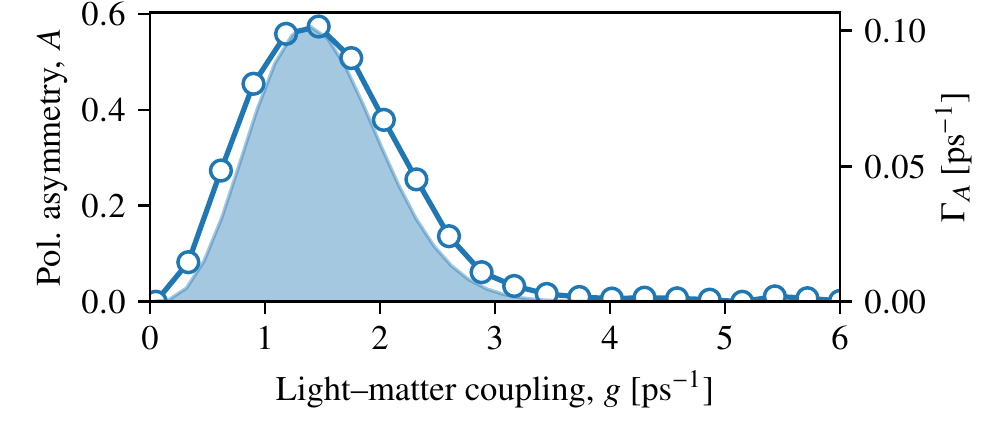}
  \caption{Asymmetry of polariton peaks, $A$, as a function of coupling strength (solid line and open circles, left axis), overlayed with analytically calculated differential polariton scattering rate, $\Gamma_A$ (shaded area, right axis). The parameters are the same as in Fig.~\ref{fig:indist} at $T=4\mathrm{\: K}$ with $\kappa=0.5\mathrm{\:ps^{-1}}$.}
  \label{fig:polariton-asymmetry}
\end{figure}

We now turn our attention towards the phonon-induced polariton asymmetry in the spectrum that arises when the upper polariton decays to the lower polariton, which is the dominant dephasing mechanism in the strong coupling regime at low temperatures. In Fig.~\ref{fig:polariton-asymmetry}, we show, as a function of $g$, the spectral asymmetry between the polariton peaks (solid line and open circles, left axis), calculated as $A=(S_--S_+)/(S_-+S_+)$, where $S_\pm := S(\omega_\pm,\omega_\pm)$ is the emission spectrum evaluated at the upper ($+$) and lower ($-$) polariton peak. As expected, the polariton symmetry is recovered in the limit $2g\gg\xi$. To support this finding, we use a master equation in the variational frame to derive the asymmetry-driving differential scattering rate from the upper to the lower polariton (see App.~\ref{sec:polariton-scattering-rates}),
\begin{align}
  \Gamma_A \simeq \frac{\pi}{2}J(2gB_\mathrm{v})[1-F^2(2gB_\mathrm{v})]
\end{align}
where $F(\nu)$ is the dimensionless variational displacement function, $F(\nu_\mathbf{k})=f_\mathbf{k}/g_\mathbf{k}$. This analytical scattering rate is also shown in Fig.~\ref{fig:polariton-asymmetry} (shaded area, right axis) and exhibits a similar behaviour as the polariton asymmetry. These findings show that the phonon-induced polariton scattering can indeed be eliminated in the phonon decoupling regime, because there are no available phonon modes with sufficiently high frequency to match the polariton energy difference. However, as shown in Figs.~\ref{fig:spectra} and \ref{fig:indist}, this does not mean that the phonons are fully decoupled in this regime, since the polaritonic phonon sidebands do not rely on resonant transitions, but occur due to vibrational dressing of the individual polaritons.

\section{Experimental platforms}
\label{sec:exper-platf}

\begin{figure}
  \centering
  \includegraphics[width=\columnwidth]{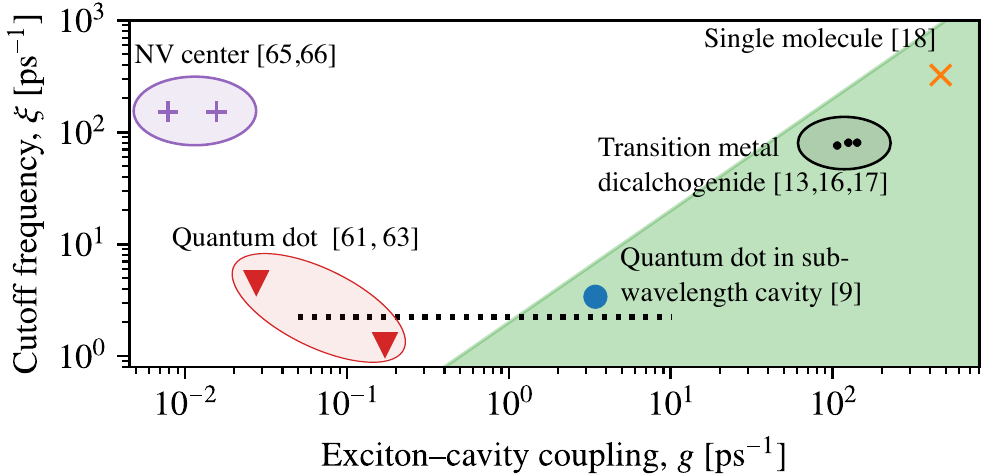}
  \caption{Examples of exciton--cavity coupling strength and phonon cutoff frequency for a range of different material platforms. The green shaded area indicates the regime $2g>\xi$, where electron--phonon decoupling can occur. The dotted line corresponds to the parameter range used in Figs.~\ref{fig:spectra} and~\ref{fig:indist}. The parameters are summarised in Table~\ref{tab:parameters}.}
  \label{fig:parameter-plot}
\end{figure}

The exciton--phonon decoupling regime can be reached by several material platforms. In Fig.~\ref{fig:parameter-plot}, we show typical values for exciton--cavity coupling strength and phonon cutoff frequency for different quantum optical systems. The green shaded area indicates the regime $2g>\xi$, where the light-matter coupling is sufficiently strong to decouple the phonons. There are several experimental examples of systems operating in this regime, namely two-dimensional transition metal dichalcogenides (black circles) and single dye molecules (orange cross) coupled to plasmonic nanocavities. In addition, we predict that recently proposed dielectric cavities with deep subwavelength confinement~\cite{choi2017self,hu2018experimental,wang2018maximizing} can bring semiconductor quantum dots into the decoupling regime (blue dot), although the current experimental state-of-the-art quantum dot cavity systems (red triangles) operate below the decoupling limit.

\section{Conclusion}
\label{sec:conclusion}

In conclusion, we have shown that the phonons in the environment of a localised exciton coupled to a nanocavity can be dynamically decoupled when the light--matter coupling is sufficiently strong. We have found that an effective decoupling occurs in the Purcell regime, where the zero-phonon transition occurs with a rate much higher than the phonon cutoff frequency. Furthermore, we have found that the phonon-induced polariton scattering in the strong light--matter coupling regime can be eliminated when the polariton splitting exceeds the phonon cutoff frequency. However, we also find a significant phonon-dressing of the individual polaritons in the high-$Q$ limit that persists into the phonon decoupling regime, demonstrating the importance of operating in the Purcell regime. These principal observations only rely on the relative magnitude of the exciton--cavity coupling strength and the phonon cutoff frequency, and generally hold for any exciton--cavity system.

\begin{acknowledgments}
The authors thank Mathias R. Jørgensen for helpful discussions.
This work was supported by the Danish National Research Foundation through NanoPhoton - Center for Nanophotonics, grant number DNRF147.
\end{acknowledgments}

\appendix
\section{Tensor network implementation}
\label{sec:tens-netw-impl}
Here we describe how the two-time correlation function $\ev{a^\dagger(t)a(t')}$ can be calculated numerically using a tensor network representation of the phonon influence functional. Our implementation of the tensor network algorithm is based on the general technique described in Ref.~\cite{jorgensen2019exploiting}, but applied to the cavity QED system investigated here. Below, we give a brief account of the method.

The relevant subspace of quantum states in the system Hilbert space are spanned by the three-dimensional basis $\{\ket{0,0},\ket{1,0},\ket{0,X}\}$; we enumerate these basis states by $\ket{s}$, where $s=0,1,2$. In general, we have $s=0,\cdots d-1$, where $d$ is the dimension of the system Hilbert space.
All dynamical information pertaining to the system degrees of freedom, comprising $n$-point temporal correlation functions, can be calculated from the so-called process tensor~\cite{pollock2018non}. This object is a rank-$2k+1$ tensor, where $k$ is the number of discrete timesteps in the numerical calculation of the dynamics. Formally, it is constructed on an enlargened system Hilbert space, where two additional copies of the system Hilbert space are created at each of the $k$ timesteps; the two partitions at timestep $t_j$ ($0\geq j \geq k$) correspond to quantum channels going into and out from, respectively, the system at time $t_j$. The infinitesimal time evolution operator, $U_{\delta t} = e^{-iH\delta t/\hbar}$, is then applied to every copy, followed by a trace over the environmental degrees of freedom. The process tensor can be formally written as
\begin{align}
\begin{split}
\Upsilon_{k:0} = &\Tr{\mathcal{U}_{\delta t}^{(s_k'r_k's_{k-1}r_{k-1})} \cdots \mathcal{U}_{\delta t}^{(s_1'r_1's_{0}r_{0})}[\chi_0^{(r_0's_0')}]} \\
&\times \dyad{s_k's_{k-1}\cdots s_1' s_0s_0'}{r_k'r_{k-1}\cdots r_1' r_0r_0'}, 
\end{split}
\end{align}
where repeated indices are summed over ($s_j,r_j = 0,1,2$) and $\mathcal{U}_{\delta t}^{(s'r'sr)}$ is a superoperator working on the environment density operator as
\begin{align}
  \mathcal{U}_{\delta t}^{(s'r'sr)}[\rho_{\rm p}] = \mel{s'}{U_{\delta t} (\dyad{s}{r}\otimes \rho_{\rm p})U_{\delta t} }{r'}
\end{align}
and $\chi_0^{(r_0's_0')}$ is the matrix element $\mel{r_0'}{\chi_0}{s_0'}$ of the initial density operator, which we take to be seperable with a thermal environmental state, $\chi_0=\rho_{\rm s}(0)\otimes e^{-\beta H_{\rm p}}/\Tr[e^{-\beta H_{\rm p}}]$ with the inverse temperature $\beta=1/k_BT$ and $\rho_{\rm s}(0)$ the initial density operator of the exciton--cavity system.
Using the process tensor, any $k$-point correlation function can be evaluated as $\ev{\mathcal{A}_k(t_k)\cdots\mathcal{A}_0(t_0)}= \Tr_{k:0}[(\mathsf{A}_k\otimes\cdots\otimes\mathsf{A}_0) \Upsilon_{k:0}]$, where the trace is over all the copies of the system Hilbert space in the process tensor and $\mathcal{A}_j$ denotes a super-operator working on one partition of the $j$th system space and $\mathsf{A}_j$ is the corresponding Choi representation. In the present situation, we are interested in evaluating two-point correlation functions $\ev{a^{\dagger}(t_{i})a(t_{i'})}$, which are obtained by setting all superoperators with $j\neq i,i'$ to the identity and $\mathcal{A}_i[\rho_{\rm s}] = a^\dagger \rho_{\rm s},\; \mathcal{A}_{i'}[\rho_{\rm s}]=a\rho_{\rm s}$.

Through the Trotter decomposition~\cite{trotter1959product}, the infinitesimal time evolution superoperator can be written as $\mathcal{U}_{\delta t} = \mathcal{V}_{\delta t/2}\mathcal{W}_{\delta t}\mathcal{V}_{\delta t/2}$, where $\mathcal{V}_{\delta t/2}$ describes free system evolution over half a timestep and $\mathcal{W}_{\delta t}$ contains the environmental dynamics and interactions, i.e. phonon scattering. In our implementation, the free evolution contains both the unitary dynamics generated by $H_{\rm s}$ and Markovian effects corresponding to cavity decay, exciton decay and temperature-dependent pure dephasing of the exciton. Formally we have $\mathcal{V}_{\delta t/2} = e^{\mathcal{L}\delta t/2}$, with the Liouvillian
\begin{align}
\begin{split}
\mathcal{L}[\rho_{\rm }] = &-\frac{i}{\hbar}[H_{\rm s},\rho_{\rm s}] + \kappa\mathcal{D}[a,\rho_{\rm s}] + \gamma\mathcal{D}[\dyad{0}{X},\rho_{\rm s}] \\ &+2\gamma^*(T)\mathcal{D}\big[\dyad{X},\rho_{\rm s}\big],
\end{split}
\end{align}
and $\mathcal{D}[A,\rho] = A\rho A^\dagger - \frac{1}{2}(A^\dagger A\rho - \rho A^\dagger A)$ is the Lindblad superoperator. The temperature-dependent pure dephasing rate is described in Sec.~\ref{sec:pure-dephasing}. In practise, we work in a frame rotating with the emitter frequency, $\omega_X$, such that $H_{\rm s}\rightarrow -\hbar\delta a^\dagger a + g(\dyad{X}{0}a + \mathrm{H.c.})$, where $\delta = \hbar \omega_X-\hbar\omega_c$. 

In turn, the Trotter decomposition allows a separation of the process tensor into a system part and an environmental influence functional, $\mathcal{F}_{k:0}$, as
\begin{align}
\Upsilon_{k:0} = \qty{\qty[\bigotimes_{j=1}^k \mathcal{V}_{\delta t/2}\otimes\mathcal{V}_{\delta t/2}^*]\mathcal{F}_{k:0}}\otimes\rho_{\rm s}(0),
\end{align}
such that all the complicated memory effects are now contained in the rank-$2k$ tensor $\mathcal{F}_{k:0}$. Importantly, this tensor has a diagonal structure, $\mathcal{F}_{k:1}^{\alpha_k \alpha_{k-1}' \cdots \alpha_1 \alpha_0'} = \hat{\mathcal{F}}_{k:0}^{\alpha_k \cdots \alpha_1}\delta_{\alpha_{k}\alpha_{k-1}'}\cdots \delta_{\alpha_1\alpha_0'}$, where $\alpha_j=(s_j,r_j)$ is a composite index and $\hat{\mathcal{F}}_{k:0}$ is a rank-$k$ tensor. This diagonal structure directly stems from the fact that $H_{\rm ep}$ is diagonal in the $\ket{s}$-basis, which is a requirement for the current formulation of the strategy to work. 

The rank-$k$ influence functional, $\hat{\mathcal{F}}_{k:0}$, can be decomposed as a product of rank-2 tensors as 
\begin{align}
\label{eq:F-decomposition}
\hat{\mathcal{F}}_{k:0}^{\alpha_k\cdots\alpha_1}=\prod_{i=1}^k\prod_{j=1}^i[b_{i-j}]^{\alpha_i\alpha_j},  
\end{align}
where $[b_{i-j}]^{\alpha_i\alpha_j}$ are the influence tensors, 
\begin{align}
    [b_{i-j}]^{\alpha_i\alpha_j} = e^{-(\lambda_{s_i}-\lambda_{r_i})(\eta_{i-j}\lambda_{s_j} - \eta_{i-j}^*\lambda_{r_j})}.
\end{align}
Here, $\lambda_s$ is the eigenvalue of $\dyad{X}$ corresponding to $\ket{s}$, i.e. $\ket{X}\ip{X}{s}=\lambda_s\ket{s}$, and $\eta_{i-j}$ are the memory kernel elements
\begin{align}
  \eta_{i-j}=
\begin{cases}
\int_{t_i-1}^{t_i}\dd{t'}\int_{t_j-1}^{t_j}\dd{t''} C(t'-t'') & i\neq j \\
\int_{t_i-1}^{t_i}\dd{t'}\int_{t_j-1}^{t'}\dd{t''} C(t'-t'') & i = j 
\end{cases}
\end{align}
with the environmental correlation function
\begin{align}
C(\tau) = \frac{1}{\pi}\int_0^\infty\dd{\nu} J(\nu)\frac{\cosh[\nu(\beta/2-it)]}{\sinh[\beta\nu/2]}.
\end{align}
The computational challenge is thus reduced to efficiently calculating the product in Eq.~\eqref{eq:F-decomposition}. In Ref.~\cite{jorgensen2019exploiting} it is described in detail how this can be carried out as the contraction of a tensor network, and the effective dimensionality of the problem can be considerably reduced through compression based on a singular value decomposition, truncating singular values below a cutoff value. As the cutoff is lowered, the numerical representation of $\hat{\mathcal{F}}_{k:0}$ converges towards the exact influence functional.

\section{Temperature-dependent pure dephasing}
\label{sec:pure-dephasing}
In the main text, we use semiconductor quantum dots for example calculations. In such systems, there is a temperature-dependent pure dephasing process that arises from virtual scattering of thermal phonons to higher-lying excitonic states~\cite{muljarov2004dephasing}. This rate is given by~\cite{reigue2017probing,tighineanu2018phonon} 
\begin{align}
\gamma^*(T) =   \frac{\alpha^2\mu}{\xi^4}\int_0^\infty\dd{\nu}\nu^{10}e^{-2(\nu/\xi)^2}n_B(\nu)[n_B(\nu)+1],
\end{align}
where $n_B(\nu)=[e^{-\nu/(k_BT)} - 1]^{-1}$ is the Bose distribution and $\mu$ together with the overall electron-phonon coupling $\alpha$ quantifies the strength of the virtual scattering process. For a typical GaAs quantum dot, we find $\mu=0.023 \:\mathrm{ps^{2}}$~\cite{denning2020phonon}, which together with $\alpha=0.025\:\mathrm{ps}^{2}$ and $\xi=2.2\:\mathrm{ps^{-1}}$ correspond to $\gamma^*(4\mathrm{\:K})=6.7\times 10^{-6}\mathrm{\:ps^{-1}}$ and $\gamma^*(150\mathrm{\: K})=0.08\mathrm{\: ps^{-1}}$.

\section{Variational minimisation of free energy}
\label{sec:variational-minimisation}
Here we derive the condition that determines the variational parameters, $f_\mathbf{k}$, by minimisation of the free energy.
In the variational frame, we write the total Hamiltonian as $H_\mathrm{v} = H_{\rm s,v} + H_{\rm e,v} + H_{\rm ep,v}$, where
\begin{align}
\label{eq:H-s-v}
\begin{split}
H_{\rm s,v} &= \hbar(\omega_{X} + R_\mathrm{v})\dyad{X} + \hbar\omega_c a^\dagger a \\ &\hspace{1.5cm} +\hbar g_\mathrm{v}(\dyad{X}{0} a + \dyad{0}{X}a^\dagger),
\end{split}
\end{align}
with $R_\mathrm{v} = \sum_\mathbf{k} f_\mathbf{k}(f_\mathbf{k}-2g_\mathbf{k})/\nu_\mathbf{k}$ the variational renormalisation of the exciton transition frequency and $g_\mathrm{v}=gB_\mathrm{v}$ is the variationally renormalised light--matter coupling strength, with the renormalisation factor $B_\mathrm{v}=\ev{e^{\pm V}}= \exp[-\frac{1}{2}\sum_\mathbf{k}\frac{f_\mathbf{k}^2}{\nu_\mathbf{k}^2}\coth(\beta\hbar\nu_\mathbf{k}/2)]$, where $\beta=1/k_{\rm B}T$.
The variational interaction Hamiltonian is given by 
\begin{align}
H_{\rm ep,v} = \hbar X B_X + \hbar YB_Y + \hbar ZB_Z, 
\end{align}
where $X=g(\dyad{X}{0}a + \dyad{0}{X}a^\dagger),\; Y=ig(\dyad{X}{0} a-\dyad{0}{X} a^\dagger),\; Z=\dyad{X}$, and $B_X=(e^{V}+e^{-V}-2B_\mathrm{v})/2,\; B_Y=i(e^{V}-e^{-V})/2,\; B_Z = \sum_\mathbf{k} (g_\mathbf{k}-f_\mathbf{k})(b_\mathbf{k}^\dagger + b_\mathbf{k})$. 
Note that the partitioning of $H_\mathrm{v}$ into system, environment and interaction terms is constructed such that $\Tr[H_{\rm ep,v}e^{-\beta H_{\rm p}}]=0$. 

The Feynman-Bogoliubov upper bound on the free energy in the variational polaron frame is~\cite{nazir2016modelling}
\begin{align}
\begin{split}
  A_B &= -\frac{1}{\beta}  \ln(\Tr[e^{-\beta H_{0,{\rm v}}}]) + \ev{H_{\rm ep,v}}_{H_{0,{\rm v}}} \\&\hspace{2cm}+ \mathcal{O}(\ev{H_{\rm ep,v}^2}_{H_{0,{\rm v}}}),
\end{split}
\end{align}
where $H_{0,{\rm v}}=H_{\rm s,v}+H_{\rm p,v}$ and $\ev*{\cdot}_{H_{0,{\rm v}}}=\Tr[\:\cdot\: e^{-\beta H_{0,{\rm v}}}]/\Tr[e^{-\beta H_{0,{\rm v}}}]$. Ignoring higher-order terms and realising that the second term vanishes by construction, we are left with the first term. 
The partition function can be factored into system and environment parts, $\Tr[e^{-\beta H_{0,{\rm v}}}] = \Tr[e^{-\beta H_{\rm p, v}}]\Tr[e^{-\beta H_{\rm s,v}}]$, where the environment part does not depend on the variational parameters, $f_\mathbf{k}$ and thus only contribute to the free energy with a constant term. The partition function of the system is given by~\cite{liu1992thermal}
\begin{align}
  \Tr[e^{-\beta H_{\rm s,v}}] = 1 + 2\sum_{n=0}^\infty \cosh(\beta\hbar\eta_{\mathrm{v},n}/2)e^{-\beta[\hbar\delta_\mathrm{v}/2 + \hbar\omega_c(n+1)]} ,
\end{align}
where $\eta_{\mathrm{v},n}=\sqrt{4g_\mathrm{v}^2(n+1) + \delta_\mathrm{v}^2},\; \delta_\mathrm{v} = \omega_{X}+R_\mathrm{v} - \omega_c:=\delta + R_\mathrm{v}$. Assuming that the thermal energy, $1/\beta$, is significantly lower than $\hbar\omega_c$, only the first term in the summation yields an appreciable contribution. 
\begin{align}
A_B \simeq -\frac{1}{\beta}\ln (1 +   2\cosh(\beta\hbar\eta_\mathrm{v}/2)e^{-\frac{1}{2}\beta\hbar(\omega_X+R_\mathrm{v} + \omega_c)}),
\end{align}
where $\eta_\mathrm{v}:=\eta_{\mathrm{v},0}$. We now require that $A_B$ is stationary with respect to $f_\mathbf{k}$, i.e. that $\partial A_B/\partial f_\mathbf{k}=0$. 
This requirement amounts to the condition
\begin{align}
\label{eq:var1}
f_\mathbf{k} = \frac{g_\mathbf{k}\qty[1 - \frac{\delta_\mathrm{v}}{\eta_\mathrm{v}}\tanh(\beta\hbar\eta_\mathrm{v}/2)]}{1 - \frac{\delta_\mathrm{v}}{\eta_\mathrm{v}}\tanh(\beta\hbar\eta_\mathrm{v}/2)\qty[1 - \frac{2g_\mathrm{v}^2}{\nu_\mathbf{k}\delta_\mathrm{v}}\coth(\beta\hbar\nu_\mathbf{k}/2)]}.
\end{align}
Since some of the quantities on the right-hand side depend on $f_\mathbf{k}$, this equation needs to be solved self-consistently.
To this end, we define the dimensionless function $F(\nu)$ such that $f_\mathbf{k} = g_\mathbf{k}F(\nu_\mathbf{k})$. Thus, we may write the renormalised quantities as 
\begin{align}
\label{eq:var2}
\begin{split}
R_\mathrm{v} &= \int_0^\infty\dd{\nu} \frac{J(\nu)}{\nu}F(\nu)[F(\nu)-2], \\
B_\mathrm{v} &= \exp[-\frac{1}{2}\int_0^\infty \dd{\nu}\frac{J(\nu)F^2(\nu)}{\nu^2}\coth(\beta\hbar\nu/2)].
\end{split}
\end{align}
Using Eqs.~\eqref{eq:var1} and \eqref{eq:var2}, the variational function $F(\nu)$ can be determined through a simple iterative numerical approach.

\section{Polariton scattering rates}
\label{sec:polariton-scattering-rates}
\begin{figure}
  \centering
  \includegraphics[width=\columnwidth]{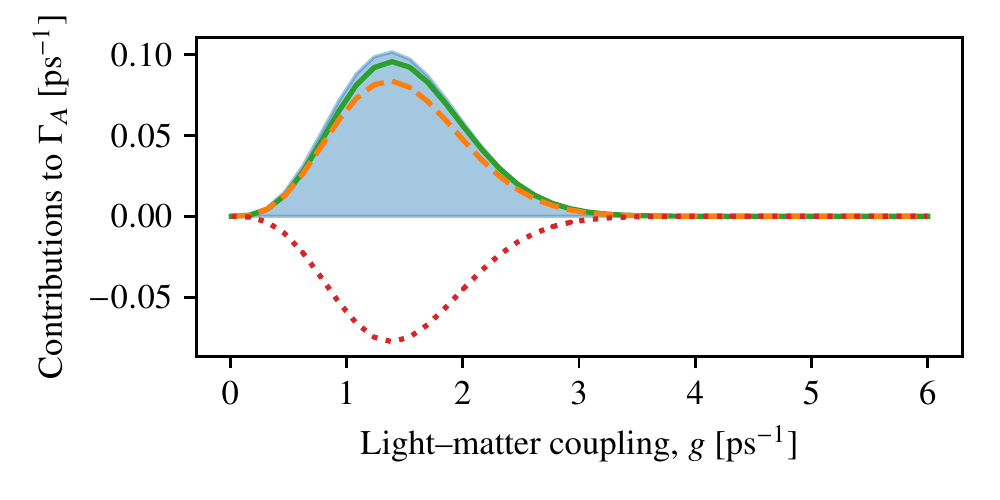}
  \caption{The three contributions to the differential polariton scattering rate, $\Gamma_A$. The shaded area shows $\Gamma_A=\epsilon_{ZZ} + \epsilon_{YY} + \epsilon_{ZY}$, the solid green line shows $\epsilon_{ZZ}$, the dashed orange line shows $\epsilon_{YY}$ and the dotted red line shows $\epsilon_{ZY}$. }
  \label{fig:polariton-asymmetry-rates}
\end{figure}

\begin{table*}[t]
  \centering
  \begin{tabular}{l|c|c}
    Material & Coupling strength, $\hbar g$ &  Cutoff frequency, $\hbar \xi$ \\
    \hline
    Transition metal dichalcogenide, $\mathrm{WS_2}$ & 93 meV~\cite{geisler2019single}, 82 meV ~\cite{qin2020revealing} & 53 meV~\cite{ramakrishna2010mos2}\\ 
    Transition metal dichalcogenide, $\mathrm{WSe_2}$ & 70 meV~\cite{kleemann2017strong} & 50 meV~\cite{del2016atypical} \\
    Single methylene blue molecule & 305 meV~\cite{chikkaraddy2016single} & 213 meV~\cite{dean2015broadband} \\
    Quantum dot in tunable microcavity & 18 $\;\mu\mathrm{eV}$~\cite{najer2019gated} & 3 meV~\cite{lobl2020radiative} \\
    Quantum dot in photonic crystal cavity & 113 $\;\mu\mathrm{eV}$~\cite{ohta2011strong} & 0.84 meV ~\cite{ota2009impact} \\
    Quantum dot in dielectric bowtie cavity & 2.0 meV~\cite{choi2017self} (th.) & 2.23 meV~\cite{denning2020phonon} \\
    NV center in photonic crystal cavity & 5 $\;\mu\mathrm{eV}$~\cite{faraon2012coupling} & 65 meV~\cite{alkauskas2014first} \\
    NV center in nanobeam photonic crystal cavity & 10 $\mu\mathrm{eV}$~\cite{mccutcheon2008design} (th.) & 65 meV~\cite{alkauskas2014first} \\ \hline
  \end{tabular}
  \caption{Light--matter coupling strength and phonon cutoff frequency for different material systems used for Fig. 5 in the main text. The annotation '(th.)' in the second column signifies coupling strengths that have been theoretically predicted.}
  \label{tab:parameters}
\end{table*}

To calculate the scattering rate between the polaritons, we start out by deriving a Markovian master equation in the variational polaron frame. Before doing so, we perform a rotating frame transformation within the exciton--cavity subspace with unitary $U=\exp[i\omega_rt(\dyad{X} + a^\dagger a)]$. Choosing the rotation frequency to be $\omega_r=\omega_{X}+R_\mathrm{v}$, the system Hamiltonian becomes
\begin{align}
  H_{\rm s,v} = -\hbar\delta_\mathrm{v} a^\dagger a + \hbar g_\mathrm{v}(\dyad{0}{X}a^\dagger + \dyad{X}{0}a)
\end{align}
The second-order Markovian master equation obtained by tracing out the phonon environment in the variational frame is then~\cite{breuer2002theory}
\begin{align}
\label{eq:variational-master-equation-1}
\begin{split}
\pdv{t}\rho(t) &= -\frac{i}{\hbar}[H_{\rm s,v},\rho] -\sum_{ij} \int_0^\infty \dd{\tau} C_{ij}(\tau) \\ &\times [A_i, \hat{A}_j(-\tau)\rho] + C_{ji}(-\tau)[\rho\hat{A}_j(-\tau),A_i],
\end{split}
\end{align}
where $A_i\in\{X,Y,Z\}$, $\hat{A}_i(-\tau)=e^{-iH_{\rm s,v}\tau}A_ie^{iH_{\rm s,v}\tau}$ is the interaction picture time evolution of the system operators and $C_{ij}(\tau)=\Tr_E[\hat{B}_i(\tau)B_j\rho_E]$. We shall use the superoperator shorthand notation $\mathcal{K}[\rho]$ to refer to the last term in Eq.~\eqref{eq:variational-master-equation-1}. From Ref.~\cite{nazir2016modelling}, we find that the correlation functions $C_{XY},\;C_{YX},\;C_{XZ}$ and $C_{ZX}$ are zero, and the remaining are given by
\begin{align}
  \begin{split}
    C_{XX}(\tau) &= \frac{B_\mathrm{v}^2}{2}(e^{\phi(\tau)} + e^{-\phi(\tau)} - 2), \\
    C_{YY}(\tau) &= \frac{B_\mathrm{v}^2}{2}(e^{\phi(\tau)} - e^{-\phi(\tau)}), \\
    C_{ZZ}(\tau) &= \int_0^\infty\dd{\nu}J(\nu)[1-F^2(\nu)] \\ &\times[\coth(\beta\hbar\nu/2)\cos(\nu\tau) -i\sin(\nu\tau)], \\
    C_{YZ}(\tau) &= -B_\mathrm{v}\int_0^\infty \dd{\nu} J(\nu)\nu^{-1}F(\nu)[1-F(\nu)]\\&\times[i\cos(\nu\tau) + \coth(\beta\nu/2)\sin(\nu\tau)], \\
    C_{ZY}(\tau) &= -C_{YZ}(\tau),
  \end{split}
\end{align}
where $\phi(\tau)=\int_0^\infty\dd{\nu}J(\nu)\nu^{-2}F^2(\nu)[\coth(\beta\hbar\nu/2)\cos(\nu\tau) -i\sin(\nu\tau)]$.
The full variational master equation (which also includes losses and dephasing as described in Sec.~\ref{sec:tens-netw-impl}), can be written on the compact form, $\pdv{t}\rho(t) = \mathcal{L}_\mathrm{v}[\rho(t)]$, where $\mathcal{L}_\mathrm{v}$ is the variational Liouvillian superoperator.
From the variational Liouvillian we can extract rates for various processes. The rate corresponding to the transition from states $\ket{\alpha}$ to $\ket{\beta}$ is given by
\begin{align}
\label{eq:transition-rates}
  \Gamma_{\alpha\beta} = \mel{\beta}{\mathcal{L}_\mathrm{v}\big[\dyad{\alpha}\big]}{\beta}
\end{align}
Here, we are particularly interested in the scattering rates between the polaritons. In the case, where the cavity and exciton are resonant in the variational frame, $\delta_\mathrm{v}=0$ (i.e. the resonance condition considered in all calculations in the main text), the polariton states within the single-excitation sector are given by $\ket{\pm} = (\ket{1,0}\pm\ket{0,X})/\sqrt{2}$. In the context of analysing the phonon-induced polariton asymmetry, we are interested in the differential polariton scattering rate, $\Gamma_A:=\Gamma_{+-}-\Gamma_{-+}$. Using the transition rate in Eq.~\eqref{eq:transition-rates}, we find that this rate has three contributions, $\Gamma_A = \epsilon_{ZZ} + \epsilon_{YY} + \epsilon_{ZY}$,
\begin{align}
\begin{split}
\epsilon_{ZZ} &= -\int_0^\infty\dd{\tau}\sin(2g_\mathrm{v}\tau)\Im{C_{ZZ}(\tau)} \\
\epsilon_{YY} &= -4g^2\int_0^\infty\dd{\tau}\sin(2g_\mathrm{v}\tau)\Im{C_{YY}(\tau)} \\
\epsilon_{ZY} &= -4g\int_0^\infty\dd{\tau}\cos(2g_\mathrm{v}\tau)\Im{C_{ZY}(\tau)}.
\end{split}
\end{align}
The first and last contributions can be calculated analytically as
\begin{align}
\begin{split}
  \epsilon_{ZZ} &= \frac{\pi}{2}J(2g_\mathrm{v})[1-F^2(2g_\mathrm{v})] \\
\epsilon_{ZY} &= \pi J(2g_\mathrm{v})F(2g_\mathrm{v})(1-F(2g_\mathrm{v})).
\end{split}
\end{align}

The contribution $\epsilon_{YY}$ cannot be resolved analytically. However, from its form, we can deduce that it vanishes in the strong-coupling limit, where $F(\nu)\rightarrow 0$ leads to $C_{YY}(\tau)\rightarrow 0$. Furthermore, in Fig.~\ref{fig:polariton-asymmetry-rates}, we show the three contributions to $\Gamma_A$ as a function of the coupling strength, showing that the contributions $\epsilon_{YY}$ and $\epsilon_{ZY}$ almost cancel out, and thus $\Gamma_A \simeq \frac{\pi}{2}J(2g_\mathrm{v})[1-F^2(2g_\mathrm{v})]$.

\section{Parameters}
\label{sec:parameters}

In Table~\ref{tab:parameters}, the parameters used for Fig. 5 in the main text are presented with references to the sources. Our theoretical prediction for a quantum dot in a dielectric bowtie cavity with deep subwavelength confinement is calculated as~\cite{andrews2015photonics}
\begin{align}
  \label{eq:1}
  g = \sqrt{\frac{d^2\omega_{eg}}{2\hbar\epsilon_0\epsilon V}},
\end{align}
where $d$ is the quantum dot dipole moment, $\epsilon$ is the dielectric constant of the background material and $V$ is the cavity mode volume. We took the dipole moment to be $d=9\times 10^{-29}\mathrm{\; Cm}$, which is a typical magnitude for self-assembled quantum dots~\cite{eliseev2000transition,muller2004determination}, the dielectric constant was taken to be 12.25, corresponding to GaAs at a wavelength of $950\mathrm{\; nm}$, which was taken as the transition wavelength. The mode volume predicted in Ref.~\cite{choi2017self} is $7.01\times 10^{-5}\lambda^3$.

For the nitrogen-vacancy (NV) center in Ref.~\cite{faraon2012coupling}, the cavity is operating in the Purcell regime. We estimated the coupling strength from the Purcell-enhanced spontaneous emission rate, $\Gamma_\mathrm{P}$, through the relation $\Gamma_\mathrm{P}=4g^2/\kappa$.

\end{document}